\documentclass[prd,nofootinbib,preprint,superscriptaddress]{revtex4}
\usepackage{graphicx}
\usepackage{multirow}
\usepackage[hypertex]{hyperref}
\newcommand{\beq}{\begin{equation}}
\newcommand{\eeq}{\end{equation}}
\newcommand{\be}{\begin{eqnarray}}
\newcommand{\ee}{\end{eqnarray}}

\newcommand{\he}{$^6$He~}
\newcommand{\neon}{$^{18}$Ne~}

\newcommand{\chr}{Cherenkov~}
\newcommand{\sig}{$3\sigma$~}

\newcommand{\ms}{\Delta m^2_{21}}
\newcommand{\ma}{\Delta m^2_{31}}

\newcommand{\sss}{\sin^2 \theta_{12}}

\newcommand{\stch}{\sin^2 2\theta_{13}}

\newcommand{\sta}{\sin^22 \theta_{23}}
\newcommand{\mst}{\Delta m^2_{21}{\mbox {~(true)}}}
\newcommand{\mat}{\Delta m^2_{31}{\mbox {~(true)}}}

\newcommand{\ssst}{\sin^2 \theta_{12}{\mbox {~(true)}}}

\newcommand{\stcht}{\sin^2 2\theta_{13}{\mbox {~(true)}}}

\newcommand{\stat}{\sin^22 \theta_{23}{\mbox {~(true)}}}
\newcommand{\dcpt}{\delta_{\mathrm{CP}}{\mbox {~(true)}}}
\newcommand{\dcp}{\delta_{\mathrm{CP}}}

\def\nue{{\nu_e}}
\def\anue{{\bar\nu_e}}
\def\numu{{\nu_{\mu}}}

\newcommand{\Sec}{Section}
\newcommand{\capdef}{}
\newcommand{\mycaption}[2][\capdef]{\renewcommand{\capdef}{#2}%
       \caption[#1]{{\footnotesize #2}}}
\makeatletter
\renewcommand{\fnum@table}{\textbf{\tablename~\thetable}}
\renewcommand{\fnum@figure}{\textbf{\figurename~\thefigure}}
\makeatother

\begin{document}
\pagestyle{plain}

\vspace*{1cm}
\begin{flushright}
\texttt{VPI-IPNAS-09-11}\\
\end{flushright}

\vspace*{1cm}
\title{Exploring neutrino parameters with a beta-beam experiment from
FNAL to DUSEL}

\author{\bf Sanjib Kumar Agarwalla}
\email{sanjib.at.vt.edu}
\affiliation{Department of Physics, Virginia Tech, Blacksburg, VA
  24061, USA}

\author{\bf Patrick Huber}
\email{pahuber.at.vt.edu}
\affiliation{Department of Physics, Virginia Tech, Blacksburg, VA
  24061, USA}

%%%%%%%%%%%%%%%%%%%%%%%%%%%%%%%%%%%%%%%%%%%%%%%%%%%%%%%%%%%%%%%%%%%%%%%%%%%%

\begin{abstract}

\vspace*{1cm}

We discuss in detail the physics reach of an experimental set-up where
electron neutrinos (anti-neutrinos) produced in a beta-beam facility
at Fermi National Accelerator Laboratory (FNAL) are sent, over a
distance of $L \simeq 1300\,\mathrm{km}$, to the Deep Underground Science and
Engineering Laboratory (DUSEL). A $300\,\mathrm{kt}$ Water \chr (WC) detector
and a $50\,\mathrm{kt}$ Liquid Argon Time Projection Chamber (LArTPC) are
considered as possible detector choices. We propose to use \neon and
\he as source ions for $\nue$ and $\anue$ beams respectively. The
maximum Lorentz boost factor, $\gamma$, available for these ions using
the Tevatron are $\gamma_{\rm Ne} = 585$ and $\gamma_{\rm He} = 350$.
This particular set-up provides the opportunity to probe the first
oscillation maximum using $\nue$ beam and the second oscillation
maximum using the $\anue$ beam which helps to evade some parameter
degeneracies. The resulting physics sensitivities for $\theta_{13}$,
CP violation and the mass hierarchy are compared to those of a
conventional superbeam from FNAL to DUSEL.

\end{abstract}
\maketitle

%%%%%%%%%%%%%%%%%%%%%%
\section{Introduction}
\label{sec:intro}
%%%%%%%%%%%%%%%%%%%%%%
The discovery of neutrino mass is one of the first clear cases of
physics beyond the Standard Model. Neutrino oscillation is now firmly
established as the leading mechanism for flavor changes in
neutrinos~\cite{solar,kl,atm,chooz,k2k,minos,limits}. This has
triggered substantial interest in precision measurements of neutrino
properties, with a special emphasis on oscillation experiments.  A new
generation of long baseline~\cite{t2k,nova} and reactor~\cite{reactor}
experiments is currently trying to measure $\theta_{13}$. This is
considered to be the first step to a large scale experimental program
of long baseline neutrino experiments which aim at determining the
neutrino mass ordering and to study leptonic CP violation, see {\it
  e.g.}~\cite{iss,issphysics}.  Due to experimental constraints on the
available neutrino flavors in both source and detector, the most
promising oscillation channels are the ones where a muon
(anti-)neutrino oscillates into an electron (anti-)neutrino or {\it
  vice versa}.  In spite of using both neutrinos and anti-neutrinos, a
serious problem with all long baseline experiments involving these
channels, arises from discrete degeneracies which manifest themselves
in three forms: the ($\theta_{13},\dcp$) intrinsic degeneracy
\cite{intrinsic}, the ($sgn(\ma),\dcp$) degeneracy \cite{minadeg}, and
the ($\theta_{23},\pi/2-\theta_{23}$) degeneracy \cite{th23octant}.
This leads to an eight-fold degeneracy \cite{eight}, with several
degenerate solutions in addition to the true one. The presence of
these degenerate solutions can severely limit the sensitivity of an
experiment.

In this paper, we will consider a beta-beam, as originally proposed
in~\cite{zucc}: a flavor pure beam of electron (anti-)neutrinos from
the beta decay of ions of short lived isotopes, which are accelerated
to a Lorentz factor, $\gamma>100$. This idea has attracted a large
community and there is a vast literature on the many different options
for such a
facility~\cite{volpe,book_betabeam,agarwalla,cernmemphys,paper1,shortnote,
optimization,pee,two_baseline,rparity,sanjib_vt1,oldpapers,donini130,doninibeta,
newdonini,bc1,bc2,fnal,betaoptim,doninialter,boulby}.
We present a study of beta-beam facility at FNAL. A somewhat similar
idea has been explored in~\cite{fnal}: there, however the emphasis was
on possible synergies between the new facility and NO$\nu$A. We, on
the other hand, focus on a full fledged, stand-alone beta-beam. We
will use \neon (for $\nue$) and \he (for $\anue$) ions.  The maximum
Lorentz boost factors using the Tevatron are $\gamma_{\rm Ne} = 585$
and $\gamma_{\rm He} = 350$ which yield energy spectra which peak at
$2.3\,\mathrm{GeV}$ and $1.4\,\mathrm{GeV}$, respectively. Our
detector is located at DUSEL~\cite{dusel,duselwhite,uslongbaseline} at
a distance of $L \simeq 1300\,\mathrm{km}$ from FNAL. We propose to
use a $300\,\mathrm{kt}$ WC detector or a $50\,\mathrm{kt}$ LArTPC as
a possible detector candidate. The first and second oscillation
maxima for the FNAL - DUSEL baseline are at $2.5\,\mathrm{GeV}$ and
$0.8\,\mathrm{GeV}$ for $\ma = 2.4 \cdot 10^{-3}\,\mathrm{eV}^2$.
Therefore, this set-up provides an unique opportunity to work at the
first oscillation maximum using the $\nue$ beam and the peak energy of
the $\anue$ beam is very close to the second oscillation maximum.

The paper is organized as follows. We begin with a brief description
of the FNAL based beta-beam facility in \Sec~2. In \Sec~3, we deal
with the relevant oscillation probabilities. In the following section
(\Sec~4) we describe the characteristics of the WC detector and the
LArTPC; also, we introduce the superbeam which will used for
comparison. We also present the expected event rates for these two
detectors. The details of our numerical technique and analysis
procedure are presented in \Sec~5. In \Sec~6, we present our results
and provide a summary.

%%%%%%%%%%%%%%%%%%%%%%%%%%%%%%%%%%%%%%%%%%%%
\section{Fermilab based Beta-beam}
\label{sec:beta-beam}
%%%%%%%%%%%%%%%%%%%%%%%%%%%%%%%%%%%%%%%%%%%%

%%%%%%%%%%%%%%%%%%%%%%%%%%%%%%%%%%%%%%%%%%%%%%%%%%%%%%%%
\begin{figure}[t]
\includegraphics[width=0.6\textwidth]{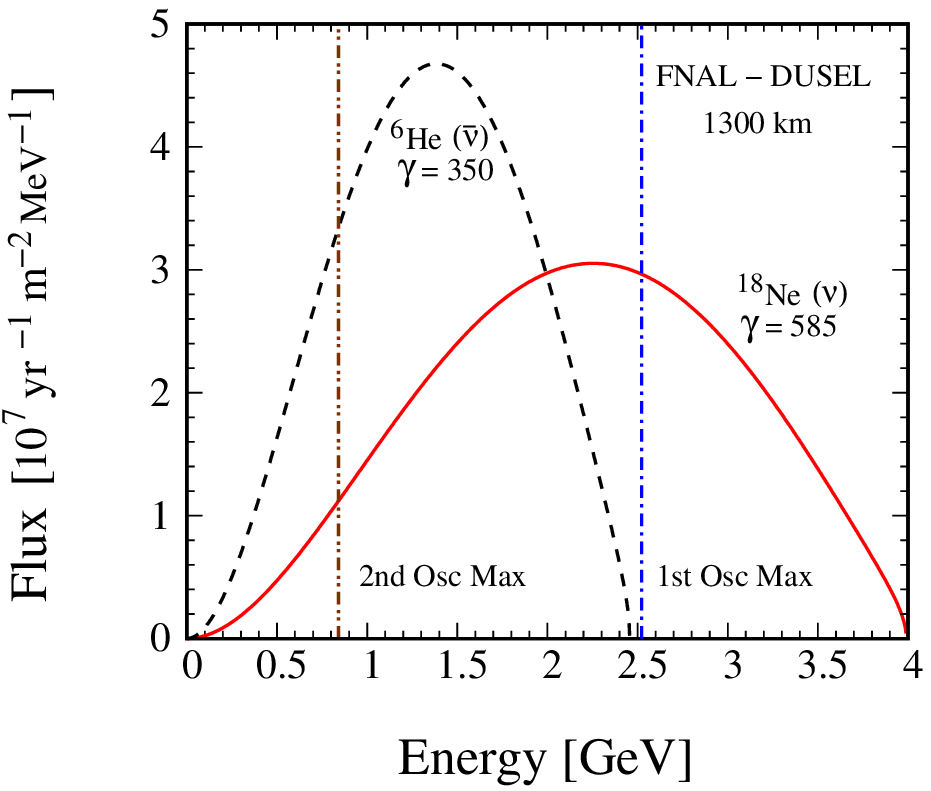}
\mycaption{\label{fig:flux} The un-oscillated beta-beam flux spectrum
  arriving at a detector placed at DUSEL at a distance of
  $1300\,\mathrm{km}$ from FNAL. The red solid line shows the
  $\nue$ spectrum generated from \neon with $\gamma = 585$. The
  black dashed curve depicts the $\anue$ spectrum originated from \he
  with $\gamma = 350$.  The blue dot-dashed and the brown
  double-dot-dashed vertical lines show the locations of first and
  second oscillation maximum for the FNAL -DUSEL baseline
  corresponding to $\ma = 2.4 \cdot 10^{-3}\,\mathrm{eV}^2$.}
\end{figure}
%%%%%%%%%%%%%%%%%%%%%%%%%%%%%%%%%%%%%%%%%%%%%%%%%%%%%%%%%

The concept of a beta-beam was proposed in~\cite{zucc}: a pure,
intense, collimated beam of electron neutrinos or their antiparticles
from the beta-decay of completely ionized unstable radioactive ions
circulating in a decay ring. The first stage in a beta-beam are
protons with an energy of a few GeV which impinge a neutron spallation
target. These neutrons, then interact in a secondary target to produce
the desired unstable isotopes, in our case \neon\ and \he (see,
table~\ref{tab:ions}). Both are noble gases and, therefore can easily
diffuse out of the secondary target, where they are collected, ionized
and bunched. Now, they can be accelerated and be put into a storage
ring with long straight sections. The decay of the highly boosted ions
in the straight sections then yields an intense, well collimated,
flavor pure electron (anti-)neutrino beam of known flux and spectrum,
see {\it e.g.}~\cite{lindroos,betabeampage}.  Feasibility of this
proposal and its physics reach is being studied in great
detail~\cite{book_betabeam}. The set-up we propose will take full
advantage of existing accelerator facilities at FNAL~\cite{fnal}. The
main limitation is the maximum energy per nucleon the Tevatron can
achieve.  The maximum $\gamma$ available for the ions considered here,
using the Tevatron as accelerator, are $\gamma_{\rm Ne} = 585$ and
$\gamma_{\rm He} = 350$. We have also studied the possibility to use
$^8$B and $^8$Li. These two isotopes have much larger endpoint
energies, of more than $10\,\mathrm{MeV}$ and thus would allow to have
a peak energy for both neutrino and anti-neutrinos around the 1st
oscillation maximum. However, since they would require a much lower
$\gamma$, they would yield a considerably smaller neutrino flux. We
have performed a full numerical study for these ions for the full
range of available $\gamma$ and found that their performance is always
worse than for \he\ and \neon. In a similar fashion, we have optimized
the $\gamma$ for \he\ and \neon\ and the values chosen above represent the
optimum in terms of physics sensitivities at the given baseline.
Since, ion production requires a proton power of only a few
$50\,\mathrm{kW}$, the currently available proton intensities may be
sufficient. A new decay ring of about $14.7\,\mathrm{km}$ in
circumference (assuming $5\,\mathrm{T}$ magnetic field and $36\%$
useful decay fraction), however, needs to be built to store the ions.

%%%%%%%%%%%%%%%%%%%%%%%%%%%%%%%%%%%%%%%%%%%%%%%%%%%
\begin{table}[t]
\begin{center}
\begin{tabular}{||c||c||c||c||c||c||c||c||} \hline \hline
   Ion & t$_{1/2}$ (s) & E$_0$ (MeV) & Total useful decays & $\gamma$ (Tevatron) & Beam & E$^{\rm peak} _{\rm lab}$ (GeV) \\
\hline
  $^{18} _{10}$Ne & 1.67 & 3.92 & $5\cdot(1.1 \cdot 10^{18})$ & 585 & $\nu_{e}$ & 2.3 \\
\hline
  $^6 _2$He & 0.81 & 4.02 & $5\cdot(2.9 \cdot 10^{18})$ & 350 & $\bar\nu_{e}$ & 1.4 \\
\hline \hline
\end{tabular}
\caption{\label{tab:ions} The beta-decay parameters, half-life
  t$_{1/2}$ and electron total end-point energy E$_0$ are shown in the
  first two columns \cite{beta}. In the third column, we list the
  total number of useful ion decays considered in this work. The
  maximum $\gamma$ available for these ions using the Tevatron are
  mentioned in column four. The peak energies of the $\nue$ and
  $\anue$ spectrum in the lab frame are shown in the last column.}
\end{center}
\end{table}
%%%%%%%%%%%%%%%%%%%%%%%%%%%%%%%%%%%%%%%%%%%%%%%%%%%%%%%%%%%%%%

While the shape of the beam spectrum depends on the end-point energy
E$_0$ and the Lorentz boost $\gamma$ of the parent ions, the flux
normalization is controlled by the number of useful ion decays per
year in one of the straight sections of the storage ring N$_{\beta}$.
Table \ref{tab:ions} depicts the relevant details of the properties of
these ions and their considered luminosities and the choices of
$\gamma$.  We have assumed $1.1 \cdot 10^{18}$ ($\nu_e$) and
$2.9\cdot 10^{18}$ ($\bar{\nu}_e$) useful ion decays per year for
\neon and \he ions respectively~\cite{beamnorm}.

Figure \ref{fig:flux} shows the un-oscillated beta-beam flux reaching
at a detector placed at the DUSEL site at a distance of
$1300\,\mathrm{km}$ from FNAL.  We see from this figure that the
$\nue$ ($\anue$) spectrum peaks at $2.3 (1.4)\,\mathrm{GeV}$. The
first and second oscillation maxima for the FNAL - DUSEL baseline are
at $2.5\,\mathrm{GeV}$ and $0.8\,\mathrm{GeV}$ for $\ma = 2.4 \cdot
10^{-3}\,\mathrm{eV}^2$.  Therefore, the $\nue$ spectrum is well
suited for the first oscillation maximum whereas the $\anue$ flux is
sensitive to the second oscillation maximum.
          
%%%%%%%%%%%%%%%%%%%%%%%%%%%%%%%%%%%%%%%%%%%%%%%%%%%%%%%%% 
\section{The $\mathbf{P_{e\mu}}$ oscillation channel}
\label{sec:prob}
%%%%%%%%%%%%%%%%%%%%%%%%%%%%%%%%%%%%%%%%%%%%%%%%%%%%%%%%%

The simulation work presented in this paper is based on the full three
flavor neutrino oscillation probabilities in matter, using the
preliminary reference Earth model for the Earth matter
density~\cite{prem}. However, to explain the nature of neutrino
oscillations as a function of baseline and/or neutrino energy, it is
crucial to use approximate analytic expression for $P_{e\mu}$ in
matter \cite{msw1,msw2,msw3}, keeping terms only up to second order in
the small quantities $\theta_{13}$ and $\alpha \equiv \ms/\ma$
\cite{golden,freund}
\be 
P_{e\mu} &\simeq&
{\underbrace{\sin^2\theta_{23} \sin^22\theta_{13}
    \frac{\sin^2[(1-\hat{A})\Delta]}{(1-\hat{A})^2} + \alpha^2
    \cos^2\theta_{23} \sin^22\theta_{12}
    \frac{\sin^2(\hat{A}\Delta)}{{\hat{A}}^2}}_{T_0}} \nonumber \\
&\pm& {\underbrace{\alpha \sin2\theta_{13} \sin2\theta_{12}
    \sin2\theta_{23} \sin(\Delta) \frac{\sin(\hat{A}\Delta)}{\hat{A}}
    \frac{\sin[(1-\hat{A})\Delta]}{(1-\hat{A})}}_{T_-}} \sin\dcp \nonumber \\
&+& {\underbrace{\alpha \sin2\theta_{13} \sin2\theta_{12}
    \sin2\theta_{23} \cos(\Delta) \frac{\sin(\hat{A}\Delta)}{\hat{A}}
    \frac{\sin[(1-\hat{A})\Delta]}{(1-\hat{A})}}_{T_+}} \cos\dcp ,
\label{eq:pemu}
\ee
where
\be
\Delta\equiv \frac{\ma L}{4E},
~~
\hat{A} \equiv \frac{A}{\ma},
~~
 A=\pm 2\sqrt{2}G_FN_eE.
\label{eq:matt}
\ee 
Here, $A$ is the matter potential, expressed in terms of the
electron density $N_e$ and the (anti)neutrino energy $E$;  `$+$'
sign refers to neutrinos whereas  `$-$' to anti-neutrinos.

%%%%%%%%%%%%%%%%%%%%%%%%%%%%%%%%%%%%%%%%%%%%%%%
\section{Event Rates at the Detectors at DUSEL}
\label{sec:event}
%%%%%%%%%%%%%%%%%%%%%%%%%%%%%%%%%%%%%%%%%%%%%%%

%%%%%%%%%%%%%%%%%
\begin{table}[t]
\begin{center}

\begin{tabular}{||c||c||c||} \hline\hline

\multicolumn{1}{||c||}{{\rule[0mm]{0mm}{6mm}\multirow{2}{*}{Detector Characteristics}}}
& \multicolumn{1}{|c||}{\rule[-3mm]{0mm}{6mm}{WC}}
& \multicolumn{1}{|c||}{\rule[-3mm]{0mm}{6mm}{LArTPC}}
\cr
& (Both $\mu^{\pm}$ \& $e^{\pm}$) & (Both $\mu^{\pm}$ \& $e^{\pm}$) \cr
& (Only QE Sample) & (QE \& IE Sample) \cr
\hline\hline
Fiducial Mass & $300\,\mathrm{kt}$ & $50\,\mathrm{kt}$ \cr\hline
Energy Threshold & WC:~A/WC:~B & 0.2 GeV \cr
\hline
Detection Efficiency ($\epsilon$) & 80\% & 80\% \cr
\hline
\multirow{2}{*}{Energy Resolution ($\delta E$) (GeV)}& \multirow{2}{*}{0.085+0.05$\sqrt{\rm E/GeV}$} 
& 0.085+0.05$\sqrt{\rm E/GeV}$ for QE Sample \cr & &  0.085+0.2$\sqrt{\rm E/GeV}$ for IE Sample \cr
\hline
Bin Size & 0.2 GeV & 0.2 GeV \cr
\hline
Background Rejection & WC:~A/WC:~B & 10$^{-3}$/10$^{-4}$ \cr
\hline
Signal error (syst.) & 2.5\% & 2.5\% \cr
\hline
Background error (syst.) & 5\% & 5\% \cr
\hline\hline
\end{tabular}
\caption{\label{tab:detector}
Detector characteristics used in the simulations. The bin size is kept fixed, 
while the number of bins is varied according to the maximum energy. We use
two different simulation methods (WC:~A \& WC:~B) to treat the backgrounds
in WC detector. Details can be found in \Sec~4.}
\end{center}
\end{table}
%%%%%%%%%%%%%%%%%%%%%%%%%%%%%%%%%%%%%%%%

Currently, two technologies for large underground neutrino detection
are considered for DUSEL: either a $300\,\mathrm{kt}$ WC detector or a
$50\,\mathrm{kt}$ LArTPC~\cite{dusel,uslongbaseline}. In the
following, we will describe their different properties as far as
needed for this work and we summarize them in
table~\ref{tab:detector}. We do not consider backgrounds due to
atmospheric neutrinos for either detector. The timing information of
the ion bunches turns out to be sufficient to reduce these backgrounds
down to an insignificant level, see, {\it e.g.} the appendix of
\cite{two_baseline}.

\subsection{Water \chr Detector}
\label{sec:wc}
%%%%%%%%%%%%%%%%%%%%%%%%%%%%%%%%%%%%%%%%%%%%%%%%%%%%%%%%%%%%%%%%%%

The water \chr technology is well understood on a large
scale~\cite{atm} and has demonstrated its excellent capability to
distinguish muons from electrons. In a beta-beam, the appearance
signal are muons from charged current $\nu_\mu$ interactions; this has
the advantage that they are easier to distinguish from neutral current
(NC) events than electrons. Nonetheless, NC events, in particular
those involving one or several neutral pions, are problematic and they
are the major source of background. Above the pion production
threshold, the background level depends on how well neutral pions can
be identified and distinguished from muons. Here, we consider only
quasi-elastic (QE) charged current events for the appearance signal.
We assume 80\% detection efficiency, $\epsilon$, for both muon and
electron QE events (see table~\ref{tab:detector}). More precisely, we
should consider single ring events, {\it i.e.} for signal events,
these events are characterized by only one charged particle being
above \chr threshold. For the NC background, this signature can be
mimicked if the two photons from a $\pi^0$ decay are so close that
the subsequent \chr rings cannot be separated. This is more likely to
happen for energetic $\pi^0$ since the opening angle between the two
photons is determined by the Lorentz $\gamma$ of the parent particle.
Note, that identifying the quasi-elastic events with the single ring
events is a reasonable approximation for the signal events. For the
spectral analysis, we include Fermi-motion by term of a constant width
of $85\,\mathrm{MeV}$~\cite{t2k} in the resolution function. The
energy resolution for the muon and electron is 5\% of $\sqrt{\rm
  E/GeV}$ and the resulting width of the Gau\ss ian energy resolution
function is the sum of both terms.

As mentioned previously, for NC background events there is no simple,
physically accurate approximation to their rate. Therefore, we will
resort to two different phenomenological parametrization. The first
one, method A, is based on an actual Super-K based Monte Carlo
simulation, whereas the second one, method B, assumes an energy
independent NC rejection efficiency. 

In method A, labeled as `WC:~A', we follow the results presented
in~\cite{ishihara} where the authors study the performance of a water
\chr detector in a beta-beam using the current simulation and analysis
tools developed for the Super-Kamiokande experiment. Results are
derived for a baseline of $700\,(130)\,\mathrm{km}$ and taking $\gamma
= 350\,(100)$ for both \neon and \he ions. As far as the background is
concerned, the main outcome of this study is, that the major
background events in the search for $\numu$ appearance signal are
indeed NC interactions involving pions. In a NC interaction, the
outgoing neutrino carries a large and generally unknown fraction of
the incoming neutrino energy. Therefore, those NC events which pass
the single ring selection criteria tend to be reconstructed with an
energy much lower than the true, incoming neutrino energy. It turns
out, that, because of this, it is possible to maximize the signal to
background ratio by imposing a cut in reconstructed energy.
Unfortunately, in reference~\cite{ishihara} only results for
$\gamma=100$ and $\gamma=350$ are presented, therefore we will
extrapolate these results to the values of $\gamma$ of interest for
this work. We assume that the average energy of a mis-reconstructed NC
event, {\it i.e.} the ones which pass the muon single ring selection,
is proportional to the incoming, true neutrino energy \emph{and} that
the proportionality constant does not change appreciable over the
energy range considered. This intuition is inspired by the form of the
differential NC cross section $d\sigma/d y$, where $y=E_q/E$, with $E$
being the energy of the incoming neutrino and $E_q$ the energy
transfered to the target. Therefore, we will assume that the
energy cut which effectively removes the NC background is proportional
to the average true neutrino event energy $\langle E \rangle_{NC}$
\beq
\langle E \rangle_{NC} = {\int \phi(E) E \sigma_{NC}(E) dE \over 
\int \phi(E) \sigma_{NC}(E) dE} \, ,
\label{eq:nc2}
\eeq
where $\phi(E)$ is the (anti-)neutrino beta-beam flux produced at the
source and $\sigma_{NC}(E)$ is the NC cross-section. We define a so
called threshold factor, $T_f$ which is different for neutrinos and
anti-neutrinos, but independent of $\gamma$
%%%%%%%%%%
\beq
E_T = T_f \langle E \rangle_{NC} \, ,
\label{eq:nc1}
\eeq
%%%%%%%%%%   
with $E_T$ being the threshold in reconstructed energy above which
there is no NC background left. It is straightforward to compute
$\langle E \rangle_{NC} = 1.57\,(1.69)\,\mathrm{GeV}$ for \neon (\he)
ions with $\gamma = 350$.  From reference~\cite{ishihara}, we find
that $E_T=1\,\mathrm{GeV}$ is sufficient for $\gamma = 350$ to
eliminate all the backgrounds for both neutrinos and anti-neutrinos.
Using equation~\ref{eq:nc1}, we obtain
%%%%%%%
\beq
T_f = 0.64\,(0.59) \quad\text{for \neon (\he)} \, .    
\label{eq:nc3}
\eeq

Reference~\cite{ishihara} provides also results for $\gamma=100$ and
we can use these to test our assumption that $T_f$ is independent of
$\gamma$.  For $\gamma = 100$, we have $\langle E \rangle_{NC} =
0.48\,(0.5)\,\mathrm{GeV}$ for \neon (\he) ions. Now using the values
of $T_f$ given by equation~\ref{eq:nc3}, we get $E_T =
0.3\,\mathrm{GeV}$ for both \neon \& \he ions which matches exactly
with the value of $E_T$ obtained in~\cite{ishihara} for $\gamma =
100$. This nicely demonstrates that the parametrization in terms of
$T_f$ and $\langle E \rangle_{NC}$ works over a reasonably large range
of energies. With $\gamma = 585$ for \neon ions we have $\langle E
\rangle_{NC} = 2.62\,\mathrm{GeV}$. Now using the value $T_f = 0.64$
given in equation~\ref{eq:nc3}, we obtain $E_T = 1.7\,\mathrm{GeV}$.
Summarizing, for method A, we use a threshold energy of
$1.7\,(1)\mathrm{GeV}$ for $\gamma = 585\,(350)$ for both neutrinos
and anti-neutrinos and assume that there is no NC background above
$E_T$.

In the second method, labeled as `WC:~B', we take a threshold of
$0.2\,\mathrm{GeV}$, which is close the production threshold for muons
and take the background to be a $10^{-3}$ of the NC current rate. The
shape of this background is identical to $\phi(E)\sigma_\mathrm{NC}(E)$
and for simplicity, we use the same energy resolution function to
smear the NC backgrounds that we use for QE signal events.

The impact of these two different simulation methods (WC:~A \& WC:~B)
is quite different while we calculate the sensitivity and we will
discuss it in detail in \Sec~6.

For our superbeam result which are shown for comparison only, we use the
identical setup as in \Sec~10 of~\cite{uslongbaseline}.

\subsection{Liquid Argon Time Projection Chamber}
%%%%%%%%%%%%%%%%%%%%%%%%%%%%%%%%%%%%%%%%%%%%%%%%%%%%

A LArTPC works due to the fact, that by applying an electric field,
free electrons created by the passage of an ionizing particle can be
drifted over large distances, $\mathcal{O}(m)$, without distortion.
This allows, to obtain three projections of the particle track by just
reading out the surface of the volume. For this reason, it seems
feasible to build very large LArTPCs at a reasonable cost. The three
projections of the track can be used to reconstruct the
three dimensional path of the particle with an accuracy of a few mm.
The currently largest LArTPC ever built, is the ICARUS T600 module~\cite{t600}
with a mass of $600\,\mathrm{t}$. It is recognized that scaling T600
by at least two orders of magnitude requires considerable R\&D.

Since this technology is still in its R\&D phase, much less knowledge
regarding its performance exists. Thus, what we present here, are
essentially educated guesses~\cite{bonnie}. We divide the signal event
into samples of QE and inelastic (IE) events because energy
reconstruction works quite differently for these two event
classes\footnote{We assume that QE and IE events are fully
  uncorrelated with each other and that they can be cleanly
  separated.}. For QE events there are only very few
tracks, typically the muon and the proton, thus it is feasible to
perform a full kinematic analysis.  Therefore, QE events will have a
rather good energy resolution. IE events, on the other hand, will
produce a large number of tracks, which the detector will not be able
to separate and reconstruct individually. Most likely, only the muon
track and a shower-like object can be identified, while the muon still
is reconstructed quite well, only summary information will be
available for the shower, {\it e.g.} total charge. Thus, the shower
energy resolution will be much worse than the muon resolution.
Therefore, we will use two different energy resolution functions for
QE and IE events~\cite{bonnie}
\begin{eqnarray}
\delta E_{QE}(E)&=&\left(0.085+0.05\sqrt{E/\mathrm{GeV}}\right)\,\mathrm{GeV}\quad\text{for QE}\,,\\
\delta E_{IE}(E)&=&\left(0.085+0.2\sqrt{E/\mathrm{GeV}}\right)\,\mathrm{GeV}\quad\text{for IE }\,.
\end{eqnarray}
 We use 80\% detection efficiency, $\epsilon$, for both muon and electron QE
and IE events (see table \ref{tab:detector}).  We calculate the
sensitivity using two different values, 10$^{-3}$ and 10$^{-4}$, for the
background rejection factor.  The present status of the simulation
study of the LArTPC at the DUSEL site can be found in \Sec~10 of
\cite{uslongbaseline}.

As we divide the signal events into two parts, we divide the NC
backgrounds into two parts as well. At a given energy $E$, the NC
backgrounds, which are relevant to estimate the sensitivity, are
calculated in the following way
\be {
\rm (NC)}^E_{total} = {\underbrace{{\rm (NC)}^E_{\rm total}
    \times \frac{\sigma_{\rm QE}}{\sigma_{\rm total}}}_{{\rm
      (NC)}^E_{\rm QE}}} + {\underbrace{{\rm (NC)}^E_{\rm total}
    \times \frac{\sigma_{\rm IE}} {\sigma_{\rm total}}}_{{\rm
      (NC)}^E_{\rm IE}}} ,
\label{eq:bkg}
\ee 
where (NC)$^E_{\rm QE}$ and (NC)$^E_{\rm IE}$ are the NC
backgrounds\footnote{The (NC)$^E_{\rm QE}$ and (NC)$^E_{\rm IE}$
  backgrounds are smeared using the same energy resolution function
  that we use for QE and IE signal events respectively. Again, this
  choice is justified more by its apparent simplicity than by its
  realism.}  applicable for QE and IE events respectively.
$\sigma_{\rm QE}$, $\sigma_{\rm IE}$ and $\sigma_{\rm total}$ are the
relevant neutrino interaction cross-sections.

%%%%%%%%%%%%%%%%%%%%%%%%
\subsection{Event Rates}
%%%%%%%%%%%%%%%%%%%%%%%%

%%%%%%%%%%%%%%%%%%%%%%%%%%%%%%%%%%%%%%%%%%%%%%%%%%%%%%%%%

The number of (anti-)muon events\footnote{In principle, both
  detectors, also are sensitive to electron (positron) events.  The
  number of electron events can be calculated using
  equation~\ref{eq:events}, by making appropriate changes to the
  oscillation probability and cross-sections.  However, as was noted
  in \cite{pee}, electron disappearance has hardly any sensitivity to
  $\theta_{13}$ and mass ordering at short baselines like the one
  under discussion. Also, it does not depend on $\dcp$.} in the $i$-th
energy bin in the detector is given by
\be 
N_{i} = \frac{T\, n_n\, \epsilon}{4\pi L^2}~ \int_0^{E_{\rm max}}
dE \int_{E_{A_i}^{\rm min}}^{E_{A_i}^{\rm max}} dE_A \,\phi(E)
\,\sigma_\numu(E) \,R(E,E_A)\, P_{e\mu}(E) \, ,
\label{eq:events}
\ee
where $T$ is the total running time, $n_n$ is the number of target
nucleons in the detector, $\epsilon$ is the detector efficiency and
$R(E,E_A)$ is the Gau\ss ian energy resolution function of the detector.
For muon (anti-muon) events, $\sigma_\numu$ is the neutrino
(anti-neutrino) interaction cross-section. The quantities $E$ and $E_A$
are the true and reconstructed (anti-)neutrino energies respectively and 
$L$ is the baseline.

%%%%%%%%%%%%%%%%%%%%%%%%%%%%%%%%%%%%%%%%%%%%%%%%%%%%%%%%
\begin{figure}[t]
\includegraphics[width=0.49\textwidth]{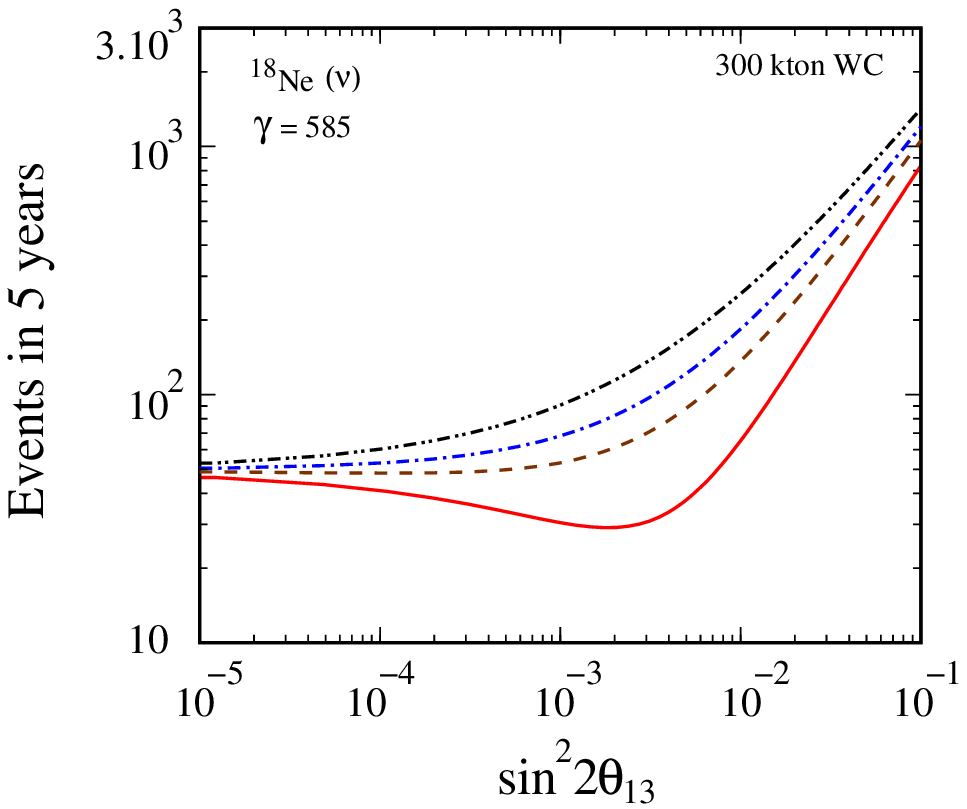}
\includegraphics[width=0.49\textwidth]{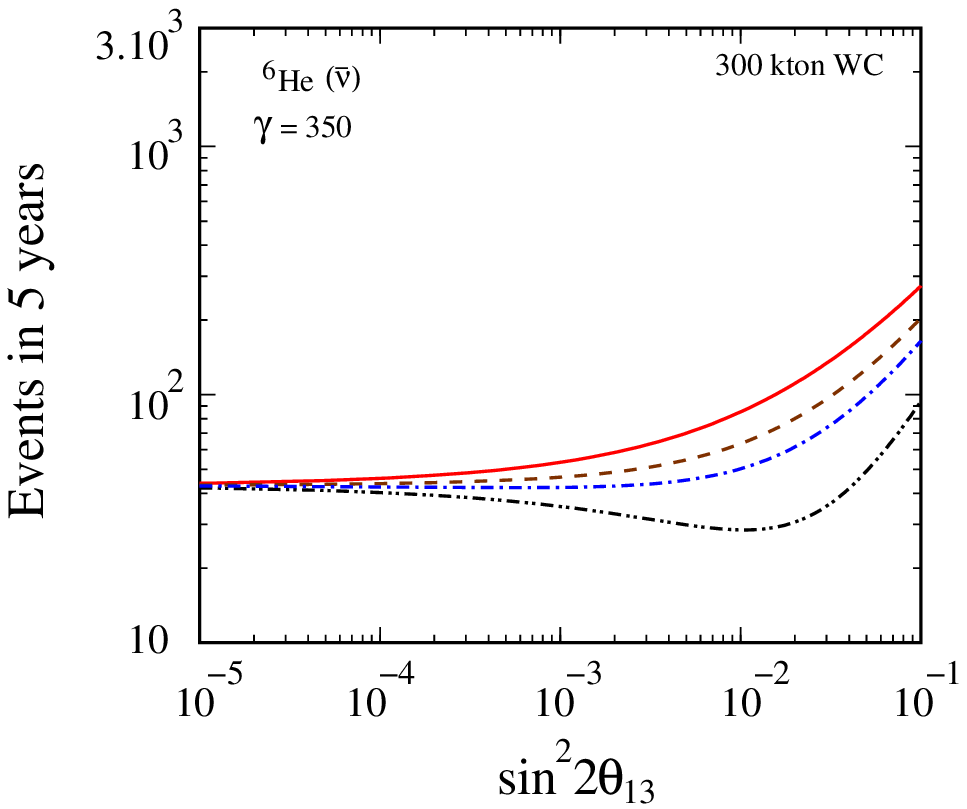}
\vskip0.5cm
\includegraphics[width=0.49\textwidth]{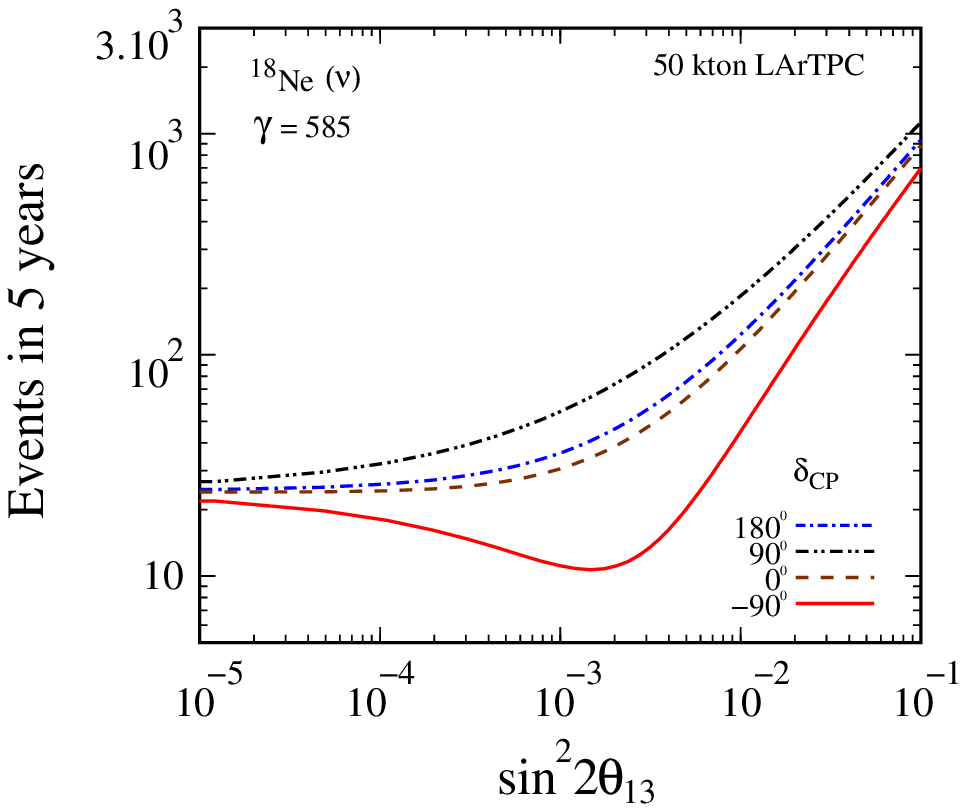}
\includegraphics[width=0.49\textwidth]{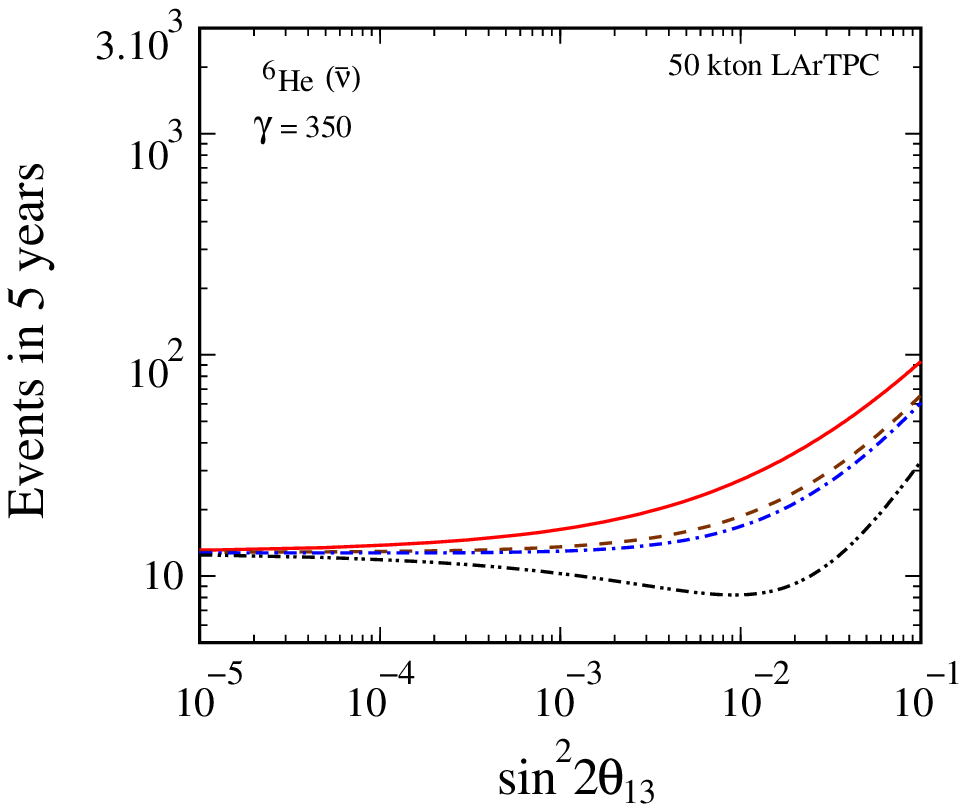}
\mycaption{\label{fig:event} Total event rates in five years as a
  function of $\stch$ in the FNAL - DUSEL set-up for \neon ($\nue$)
  with $\gamma = 585$ and \he ($\anue$) with $\gamma = 350$ are shown.
  The upper panels are for a $300\,\mathrm{kt}$ WC detector, while the
  lower ones are for a $50\,\mathrm{kt}$ LArTPC.  Results are
  depicted for four different values of $\dcp$: -90$^\circ$,
  0$^\circ$, 90$^\circ$, and 180$^\circ$.  Normal hierarchy has been
  assumed. For all other oscillation parameters we use the
 values given in equation~\ref{eq:central}.}
\end{figure}
%%%%%%%%%%%%%%%%%%%%%%%%%%%%%%%%%%%%%%%%%%%%%%%%%%%%%%%%%

We simulate the signal event spectrum using equation~\ref{eq:events}
for our assumed true\footnote{We distinguish between the ``true''
  values of the oscillation parameters, which are used to compute the
  data, and their fitted values. Throughout this paper we denote the
  true value of a parameter by putting ``(true)'' after the symbol for
  the parameter.} values for the set of oscillation parameters as
given in equation~\ref{eq:central}. The left-hand panels of
figure~\ref{fig:event} portray the number of neutrino events expected
at DUSEL from five years exposure of the \neon beta-beam from FNAL
with $\gamma = 585$ while the right-hand panels show the anti-neutrino
events from five years run of the \he beta-beam with $\gamma = 350$.
Results are presented as a function of $\stch$ for four different
values of $\dcp$: -90$^\circ$, 0$^\circ$, 90$^\circ$, and 180$^\circ$
with normal mass hierarchy.  The upper panels are for a
$300\,\mathrm{kt}$ WC detector, while the lower panels are for a
$50\,\mathrm{kt}$ LArTPC.  The number of events (both for \neon and
\he) varies in a wide range with the choice of $\dcp$. Out of the four
different choices of $\dcp$, the maximum (minimum) number of events
for neutrinos is obtained with 90$^\circ$ (-90$^\circ$) irrespective
of the choice of $\stch$.  For anti-neutrinos, the same is true with
$\dcp\rightarrow -\dcp$.  We can explain this fact with the help of
equation~\ref{eq:pemu}. At the FNAL - DUSEL baseline, matter effects
are small and hence $\hat{A}\ll 1$ in equation~\ref{eq:matt}.  The
oscillation probabilities can be expressed as $P_{e\mu} \simeq T_0 +
T_- \sin \dcp + T_+ \cos \dcp$ and $P_{\bar{e}\bar{\mu}} \simeq T_0 -
T_- \sin \dcp + T_+ \cos \dcp$, where $T_0, T_\pm$ are independent of
$\dcp$, whence the symmetry is manifest.  For most of the energies
$T_-$ is positive assuming normal hierarchy.  Now for $\dcp=90^\circ$
(-90$^\circ$) the term $T_- \sin \dcp$ gives a positive (negative)
contribution towards the probability for neutrinos. The opposite is
true for anti-neutrinos. The behavior of the number of events with
$\sin^2 2\theta_{13}$ is also understandable. As we increase the value
of $\stch$ from a near-zero value, the 2nd and 3rd terms in
equation~\ref{eq:pemu}, which are linear in $\sin 2\theta_{13}$ and
are dependent on $\dcp$, begin to contribute at first. Beyond a
certain value of $\stch$, the first term, which does not depend on
$\dcp$, takes over leading to the rise for all curves, but the
relative size of CP effects decreases. This is the reason, why the
discovery of leptonic CP violation does not become any easy task at
large $\stch$.

As far as the total number of neutrino events is concerned, the
performance of the different detectors is less different than
indicated by a factor of six difference in the detector masses (see
the upper and lower left panels of figure~\ref{fig:event}). The reason
is, that for the WC detector we consider only QE events but for the
LArTPC we take into account both QE and IE events.  At the same time
the IE cross section is larger than the QE one at the relevant
energies, see figure~\ref{fig:flux}.

\subsection{Reference set-up}
\label{sec:wbb}

In order to compare the beta-beam set-up to alternative possibilities
at FNAL, we introduce the wide-band beam concept. This project has
been studied in detail in~\cite{wbb,uslongbaseline}. Here, a
conventional neutrino beam will be sent from FNAL to a
$300\,\mathrm{kt}$ WC in DUSEL. This beam has a spectrum wide enough
to cover the first and second oscillation maximum and therefore can
resolve most degeneracies. We use the same implementation as
in~\cite{wbb}, with exception of the beam spectrum, which has been
updated to correctly represent the beam which would be produced at
FNAL~\cite{Bishai:2008zza}. Also, the beam intensity now corresponds
to $1.2\,\mathrm{MW}$, this is the maximum beam power which can be
achieved at FNAL without Project~X~\cite{fnalprotonplan}. The
resulting physics sensitivities have been computed using {\sf
  GLoBES}~\cite{globes} and are shown as green shaded regions in
figure~\ref{fig:sensitivity}.

%%%%%%%%%%%%%%%%%%%%%%%%%%%%%%%%%
\section{The Numerical Technique}
\label{sec:numerics}
%%%%%%%%%%%%%%%%%%%%%%%%%%%%%%%%%

For all calculations we use as central (true) values
\begin{eqnarray}
\label{eq:central}
\left|\Delta m^2_{31}\right|=2.4\cdot10^{-3}\,\mathrm{eV}^2 \pm 5\%\,,&\quad&\sin^22\theta_{23}=1.0\pm 1\%\,\nonumber\\
\Delta  m^2_{21}=7.6\cdot10^{-3}\,\mathrm{eV}^2 \pm 2\%\,,&\quad&
\sin^2\theta_{12}=0.32\pm 6\%\,.
\end{eqnarray}
In all fits, these parameters are allowed to vary within the stated
$1\,\sigma$ intervals. The central values are the current best fit
values~\cite{Maltoni:2004ei}; also the errors on the solar parameters
are taken from~\cite{Maltoni:2004ei}. The errors on the atmospheric
parameters correspond to the result which are expected from T2K and
NO$\nu$A~\cite{Huber:2009cw}.

Here, we describe in detail the numerical procedure adopted to
calculate the discovery potential of the FNAL - DUSEL beta-beam
set-up.  For our statistical analysis we use the following $\chi^2$
functions for WC detector and LArTPC \be (\chi^2_{total})_{\rm WC} &=&
\chi^2_{(\nue \rightarrow \numu)_{\rm QE}}
+ \chi^2_{(\bar\nue \rightarrow \bar\numu)_{\rm QE}} \nonumber \\
&+& \chi^2_{(\nue \rightarrow \nue)_{\rm QE}}
+ \chi^2_{(\bar\nue \rightarrow \bar\nue)_{\rm QE}} \nonumber \\
&+& \chi^2_{prior}
\label{eq:tot_chisq_wc}
\ee 
and 
\be
(\chi^2_{total})_{\rm LArTPC} &=& \chi^2_{(\nue \rightarrow \numu)_{\rm QE}}
               + \chi^2_{(\nue \rightarrow \numu)_{\rm IE}}
               + \chi^2_{(\bar\nu_e \rightarrow \bar\numu)_{\rm QE}}
               + \chi^2_{(\bar\nu_e \rightarrow \bar\numu)_{\rm IE}} \nonumber \\
              &+& \chi^2_{(\nu_e \rightarrow \nue)_{\rm QE}}
               + \chi^2_{(\nu_e \rightarrow \nue)_{\rm IE}}
               + \chi^2_{(\bar\nu_e \rightarrow \bar\nue)_{\rm QE}}
               + \chi^2_{(\bar\nu_e \rightarrow \bar\nue)_{\rm IE}} \nonumber \\
              &+& \chi^2_{prior}~.
\label{eq:tot_chisq_LArTPC}
\ee
The $\chi^2_{(\nue \rightarrow \numu)_{\rm QE}}$ is given by
\be
\chi^2_{(\nue \rightarrow \numu)_{\rm QE}} = min_{\xi_s, \xi_b}\left[2\sum^{n}_{i=1}
(\tilde{y}_{i}-x_{i} - x_{i} \ln \frac{\tilde{y}_{i}}{x_{i}}) +
\xi_s^2 + \xi_b^2\right ]~,
\label{eq:chipull}
\ee
where $n$ is the total number of bins and
\be
\tilde{y}_{i}(\{\omega\},\{\xi_s, \xi_b\}) = N^{th}_i(\{\omega\}) \left[
1+ \pi^s \xi_s \right] +
N^{b}_i \left[1+ \pi^b \xi_b \right]~.
\label{eq:rth}
\ee
Above, $N^{th}_i(\{\omega\})$ is the predicted number of QE events
(calculated using equation~\ref{eq:events}) in the $i$-th energy bin
for a set of oscillation parameters $\omega$ and $N_i^b$ are the
number of background events in bin $i$. The quantities $\pi^s$ and
$\pi^b$ in equation~\ref{eq:rth} are the systematical errors on
signals and backgrounds respectively. We consider $\pi^s = 2.5\%$ and
$\pi^b = 5\%$ (see table~\ref{tab:detector}). The quantities $\xi_s$
and $\xi_b$ are the pulls due to the systematical error on signal and
background respectively. The data from equation~\ref{eq:chipull}
enters through the variable $x_i=N_i^{ex}+N_i^b$, where $ N_i^{ex}$ is
the number of observed QE signal events in the detector and $N_i^b$ is
the background, as mentioned earlier.  We simulate the QE signal event
spectrum using equation~\ref{eq:events} for our assumed true values
for the set of oscillation parameters given in
equation~\ref{eq:central}. We consider all the values of $\stcht$ and
$\dcpt$ in their allowed range and assume NH as true hierarchy.  In a
similar way, we estimate the contributions towards $\chi^2_{total}$
coming from other oscillation channels and event types (for both
neutrino and anti-neutrino modes).  In our $\chi^2$ fit we marginalize
over {\it all} oscillation parameters and as well as the neutrino mass
ordering, as applicable. We perform this by allowing all of these to
vary freely in the fit and picking the smallest value of the $\chi^2$
function. However, we assume that some of these parameters which are
poorly constrained by this experimental set-up, will be measured
better from other experiments.  Therefore, we impose a prior, or
external constraint, on these parameters through $\chi^2_{prior}$,
given by
\be
\chi^2_{prior} &=& \left (\frac{|\Delta m^2_{31}|-
    |\mat|}{\sigma(|\Delta m^2_{31}|)} \right )^2 +
\left (\frac{\sta-\stat}{\sigma(\sta)} \right )^2\nonumber \\
&+& \left (\frac{\Delta m^2_{21}- \mst}{\sigma(\Delta m^2_{21})}
\right )^2 + \left (\frac{\sss-\ssst}{\sigma(\sss)} \right )^2~.
\label{eq:prior}
\ee
where the $1\sigma$ errors on these, are given in
equation~\ref{eq:central}.  We minimize the $\chi^2_{total}$ using the
same procedure as it was described in the appendix of
\cite{shortnote}.
    
%%%%%%%%%%%%%%%%%%%%%%%%%%%%%%%%%%%%%%%%%%%%%%%%%%%
\section{Results \& Summary}
\label{sec:results}
%%%%%%%%%%%%%%%%%%%%%%%%%%%%%%%%%%%%%%%%%%%%%%%%%%%%

%%%%%%%%%%%%%%%%%%%%%%%%%%%%%%%%%%%%%%%%%%%%%%%%%%%%%%%%
\begin{figure}[t]
\includegraphics[width=0.6\textwidth]{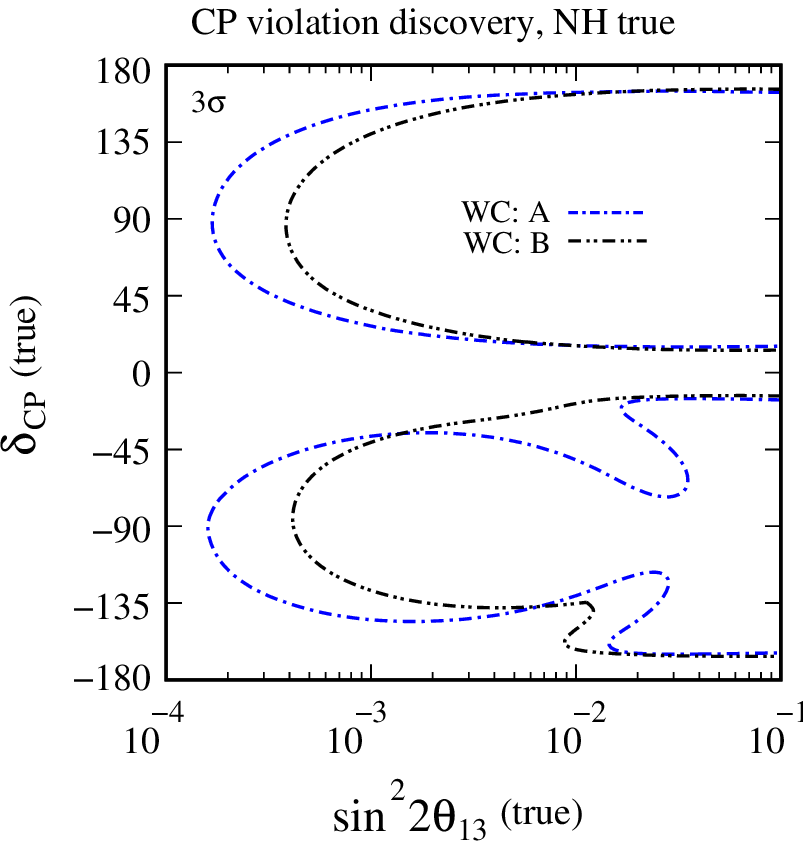}
\mycaption{\label{fig:730km} CP violation discovery potential of beta-beam 
at $730\,\mathrm{km}$ as range of $\dcpt$ as a function of the $\stcht$
assuming normal hierarchy as true hierarchy at the 1 d.o.f. \sig C.L.
Here we take $\gamma = 350$ for both \neon \& \he ions. The results are shown
for $300\,\mathrm{kt}$ WC detector assuming two different simulation methods.
See the text for details.}
\end{figure}
%%%%%%%%%%%%%%%%%%%%%%%%%%%%%%%%%%%%%%%%%%%%%%%%%%%%%%%%%

We evaluate the physics reach of the FNAL - DUSEL beta-beam set-up in
terms of its discovery potentials for $\stch$, CP violation and the
mass hierarchy.  These discovery potentials quantify for any given
$\stcht$ for which range of possible values of $\dcpt$ the
corresponding quantity will be discovered or measured at the chosen
confidence level. The discovery reach for $\stch$ is defined by the
minimum value of $\stcht$ which allows us to rule out $\stch=0$ in the
fit. The CP violation discovery potential is defined as the range of
$\dcpt$ as a function of $\stcht$ for which one can use the data to
exclude the CP conserving solutions $\dcp=0^\circ$ and
$\dcp=180^\circ$.  The mass hierarchy discovery reach is the limiting
value of $\stcht$ for which the wrong hierarchy can be excluded.

Before we present our results for FNAL - DUSEL beta-beam set-up, we
would like to discuss the CP violation discovery potential of
beta-beam using a $300\,\mathrm{kt}$ WC detector at $730\,\mathrm{km}$
taking $\gamma = 350$ for both \neon \& \he ions (see
figure~\ref{fig:730km}). This setup has already been considered in the
literature and has been shown provide high performance~\cite{bc1}.
The first and second oscillation maxima for $730\,\mathrm{km}$
baseline are at $1.4\,\mathrm{GeV}$ and $0.5\,\mathrm{GeV}$ for $\ma =
2.4 \cdot 10^{-3}\,\mathrm{eV}^2$.  At this baseline, both the $\nue$
and the $\anue$ beam peak at the first oscillation maximum for $\gamma
= 350$. Also for this setup, we use two methods to treat the NC
background in the water Cherenkov detector as described in
section~\ref{sec:wc}. For method A, we have a threshold energy of
$1\,\mathrm{GeV}$ for both \neon and \he ions and assume no
backgrounds above the threshold. Since there is no background left
above $1\,\mathrm{GeV}$, it is possible to probe CP violation at very
small values of $\stcht$ as depicted by the dash-dotted blue line of
figure~\ref{fig:730km}. The major drawback of this method is that we
do not have any access to the second oscillation maximum. Therefore,
the ($sgn(\ma),\dcp$) degeneracy for large values of $\stcht$ is fully
developed; as a result there is a sizable gap in sensitivity around
$\stcht = 3 \cdot 10^{-2}$. This effect has been described
in~\cite{Huber:2002mx} and has been termed $\pi$-transit.

For method B, shown as dash-double-dotted black line in
figure~\ref{fig:730km}, where the threshold is $0.2\,\mathrm{GeV}$ and
a background rejection factor of 10$^{-3}$ is used for both the ions.
Although the total backgrounds are higher, the second oscillation
maximum can be used, nonetheless. Therefore, the effects of
$\pi$-transit are mitigated for most of the parameter space, {\it
  i.e.} the sensitivity at large $\theta_{13}$ has essentially no
gaps.  The higher backgrounds result, however, in a reduced reach for
small values of $\theta_{13}$. This concludes our comparison with
previous results.

%%%%%%%%%%%%%%%%%%%%%%%%%%%%%%%%%%%%%%%%%%%%%%%%%%%%%%%%
\begin{figure}[t]
\includegraphics[width=0.328\textwidth]{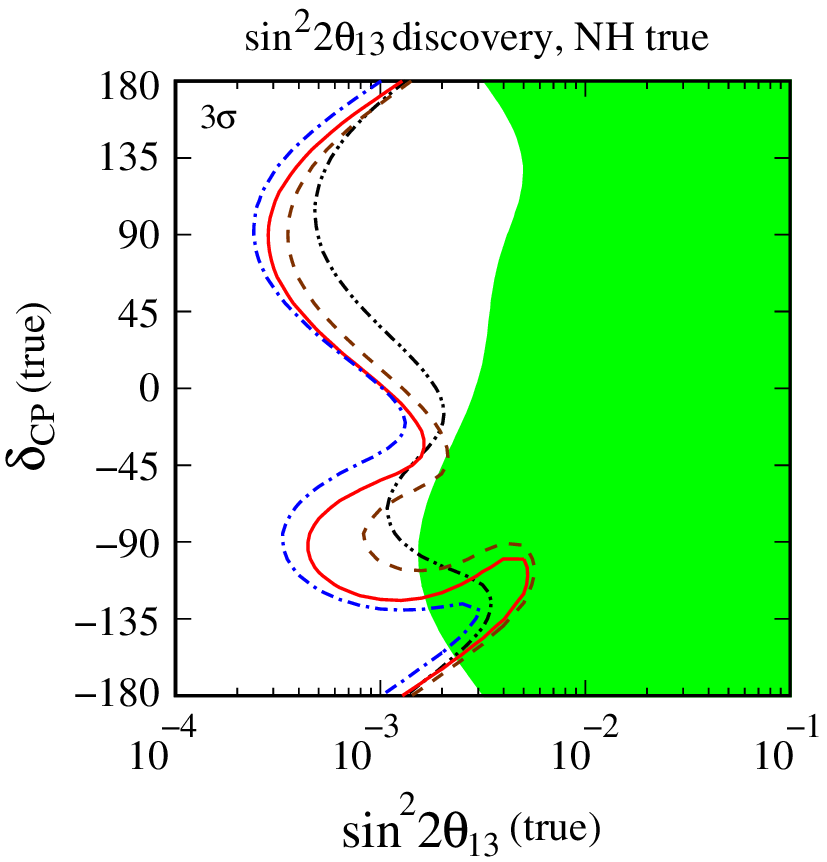}
\includegraphics[width=0.328\textwidth]{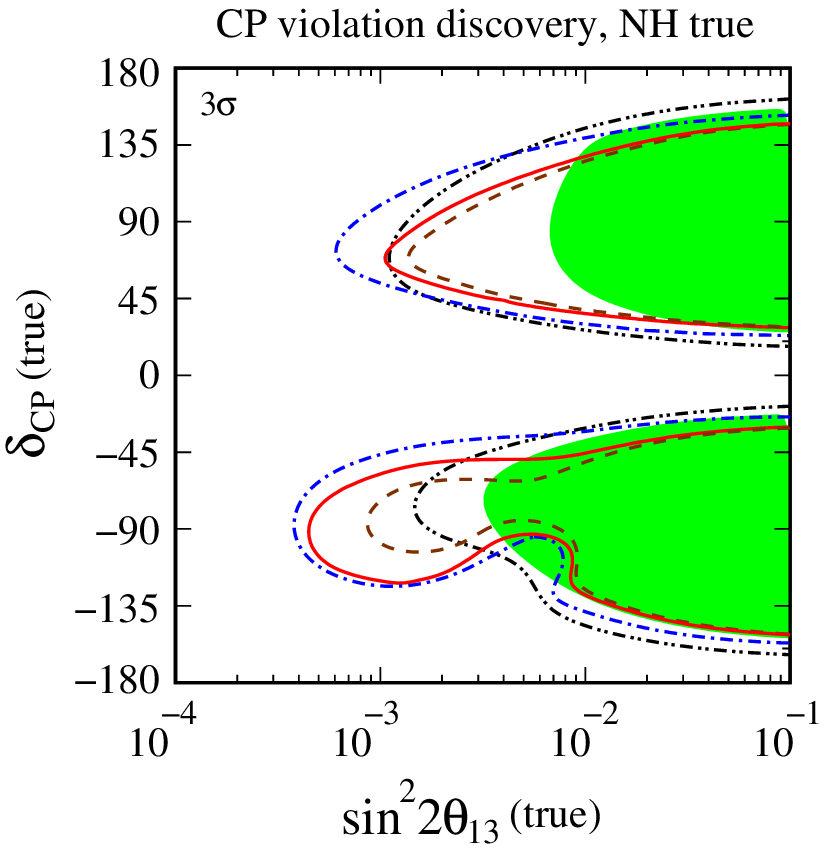}
\includegraphics[width=0.328\textwidth]{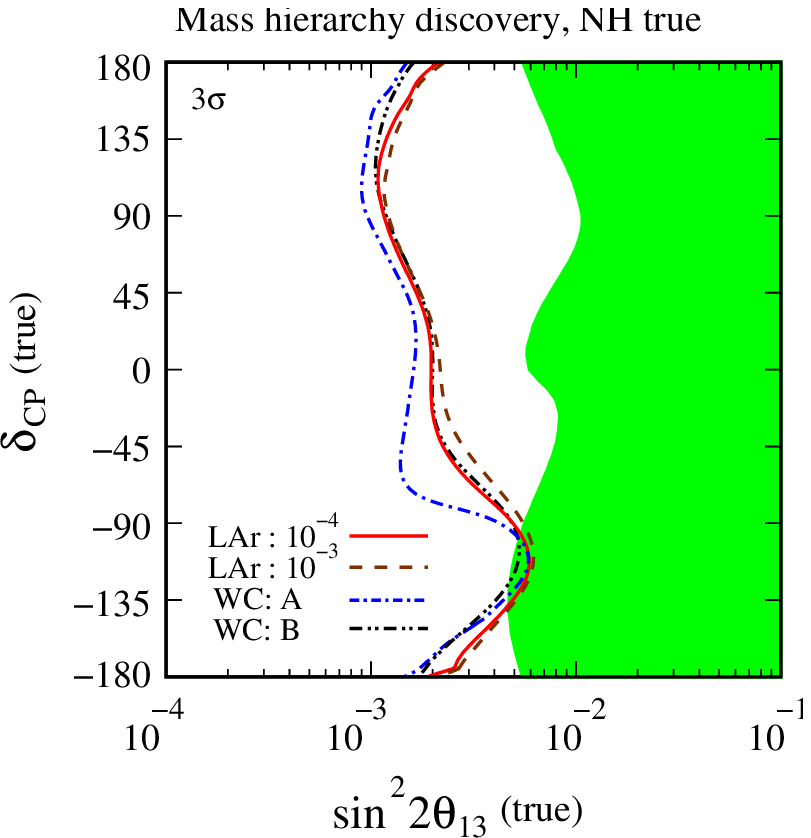}
\mycaption{\label{fig:sensitivity} Performance of FNAL - DUSEL
  beta-beam set-up at the 1 d.o.f. \sig C.L. in addressing $\stch$, CP
  violation, and mass hierarchy discovery potential as range of
  $\dcpt$ as a function of the $\stcht$ assuming normal hierarchy as
  true hierarchy. The results are shown for $300\,\mathrm{kt}$ WC
  detector (assuming the simulation methods WC:~A \& WC:~B. See
  \Sec~4 for details) and $50\,\mathrm{kt}$ LArTPC (assuming the
  background rejection factor of 10$^{-3}$ and 10$^{-4}$). The green
  shaded regions show the sensitivity of a wide band beam using the WC
  detector, as defined in detail in section~\ref{sec:wbb}.}
\end{figure}
%%%%%%%%%%%%%%%%%%%%%%%%%%%%%%%%%%%%%%%%%%%%%%%%%%%%%%%%%%%%%%%%%%%%%%%

Our results are summarized in figure~\ref{fig:sensitivity}, where the
physics sensitivities for $\stch$ (left panel), CP violation (middle
panel), and for the mass hierarchy (right panel) are shown for a
baseline of $1300\,\mathrm{km}$. Results are shown at the
$3\,\sigma$ confidence level for one degree of freedom. The various
line styles are for the different detector options and background
levels as given in the legend. The green shaded areas are the
corresponding results for a conventional superbeam from FNAL to DUSEL
using a $300\,\mathrm{kt}$ WC detector as described in detail
in~\cite{wbb}.

For the WC detector, we observe a distinct difference in sensitivities
between the two different schemes to include background (black,
dash-double-dotted line and blue, dash-dotted line). Clearly, using a
hard energy cut (WC:~A, blue, dash-dotted line), which eliminates
essentially all background and all information from the second
oscillation maximum, performs better than WC:~B (black,
dash-double-dotted line).  The only exception is found for the CP
violation reach around $\sin^22\theta_{13}=5\cdot10^{-3}$ for a CP
phase of approximately $-120^\circ$. The reason, is the loss in
information from the second oscillation maximum. Taking WC:~A as our
benchmark case, we find that a beta beam would result in about one
order of magnitude improvement over a superbeam for all three
measurements for at least one half of the CP phases.  The performance
of a $50\,\mathrm{kt}$ LArTPC is rather similar to the one of a six
times larger WC detector using background scheme B.

In comparison to the superbeam, shown as green shaded regions and
described in section~\ref{sec:wbb}, we see that using the same
$300\,\mathrm{kt}$ WC detector the gain from using a beta-beam is
obvious and most pronounced for $0<\delta_{CP}<180^\circ$. This
statement holds also for the use of a six times smaller
$50\,\mathrm{kt}$ LArTPC. It is left to the judgement of the reader,
whether the physics gain from a beta beam is commensurate to the
effort of building a decay ring and of running the Tevatron for
another decade. We also have shown that, given the performance
parameters of the Tevatron, a baseline of around $730\,\mathrm{km}$
would provide a much more compelling physics gain.  However, this
baseline violates the boundary condition of locating the far detector
at DUSEL. In summary, while a beta beam is a very interesting option
to pursue high precision neutrino physics, especially the search for
CP violation, it seems not to fit very well with existing and planed
infrastructure in the US.

%%%%%%%%%%%%%%%%%%%%%%%%%%%%%%%%
\acknowledgments

We would like to thank J.~Link and D.~Mohapatra for useful
discussions. In particular we thank M.~Mezzetto for detailed
information on backgrounds in a Water Cherenkov detector. We
acknowledge support from the U.S. Department of Energy under award
number DE-SC0003915.

%%%%%%%%%%%%%%%%%%%%%%%%%%%%%%%%%%%%%%%%%%%%%%%%%%%%%%%%%%%%%%%%


\begin{thebibliography}{99}

\bibitem{solar}
B.~T.~Cleveland {\it et al.},
Astrophys.\ J.\  {\bf 496}, 505 (1998);
%%CITATION = ASJOA,496,505;%%
%
J.~N.~Abdurashitov {\it et al.}  [SAGE Collaboration],
J.\ Exp.\ Theor.\ Phys.\  {\bf 95}, 181 (2002)[Zh.\ Eksp.\ Teor.\ Fiz.\  {\bf 122}, 211 (2002)];
%%CITATION = ZETFA,122,211;%%
%
W.~Hampel {\it et al.}  [GALLEX Collaboration],
Phys.\ Lett.\ B {\bf 447}, 127 (1999);
%%CITATION = PHLTA,B447,127;%%
%
S.~Fukuda {\it et al.}  [Super-Kamiokande Collaboration],
Phys.\ Lett.\ B {\bf 539}, 179 (2002);
%%CITATION = PHLTA,B539,179;%%
%
B.~Aharmim {\it et al.}  [SNO Collaboration],
Phys.\ Rev.\ C {\bf 72}, 055502 (2005);
%%CITATION = PHRVA,C72,055502;%%
%
C.~Arpesella {\it et al.}
  [Borexino~Collaboration],
  %``First real time detection of Be7 solar neutrinos by Borexino,''
  Phys.\ Lett.\  B {\bf 658}, 101 (2008).
  %%CITATION = PHLTA,B658,101;%%

\bibitem{kl}
T.~Araki {\it et al.}  [KamLAND Collaboration],
Phys.\ Rev.\ Lett.\  {\bf 94}, 081801 (2005);
%%CITATION = PRLTA,94,081801;%%
%
  S.~Abe {\it et al.}  [KamLAND Collaboration],
  %``Precision Measurement of Neutrino Oscillation Parameters with KamLAND,''
  arXiv:0801.4589 [hep-ex].
  %%CITATION = ARXIV:0801.4589;%%

\bibitem{atm}
  Y.~Fukuda {\it et al.}  [Super-Kamiokande Collaboration],
  %%CITATION = PHRVA,D75,097302;%%
  %``Evidence for oscillation of atmospheric neutrinos,''
  Phys.\ Rev.\ Lett.\  {\bf 81}, 1562 (1998);
%  [arXiv:hep-ex/9807003].
  %%CITATION = PRLTA,81,1562;%%
  Y.~Ashie {\it et al.}  [Super-Kamiokande Collaboration],
%   ``A measurement of atmospheric neutrino oscillation parameters by
  %Super-Kamiokande I,''
  Phys.\ Rev.\ D {\bf 71}, 112005 (2005).
%  [arXiv:hep-ex/0501064].
%%CITATION = PHRVA,D71,112005;%%

\bibitem{chooz}
M.~Apollonio {\it et al.},
%``Search for neutrino oscillations on a long base-line at the CHOOZ  nuclear
%power station,''
Eur.\ Phys.\ J.\ C {\bf 27}, 331 (2003).
%%CITATION = EPHJA,C27,331;%%

\bibitem{k2k}
E.~Aliu {\it et al.}  [K2K Collaboration],
  %``Evidence for muon neutrino oscillation in an accelerator-based
  %experiment,''
  Phys.\ Rev.\ Lett.\  {\bf 94}, 081802 (2005).

\bibitem{minos}
D. G. Michael {\it et al.}, [MINOS Collaboration],
%   ``Observation of muon neutrino disappearance with the {MINOS} detectors and
  %the {NuMI} neutrino beam,''
  arXiv:hep-ex/0607088.
  %%CITATION = HEP-EX 0607088;%%

\bibitem{limits}
  M.~C.~Gonzalez-Garcia and M.~Maltoni,
  %``Phenomenology with Massive Neutrinos,''
 Phys.\ Rept.\  {\bf 460}, 1 (2008);
  %arXiv:0704.1800 [hep-ph];
  %%CITATION = ARXIV:0704.1800;%%
%
  S.~Choubey,
%   ``Probing The Neutrino Mass Matrix In Next Generation Neutrino Oscillation
  %Experiments,''
  arXiv:hep-ph/0509217;
  %%CITATION = PANUE,69,1930;%%
%
  S.~Goswami,
  %``Neutrino oscillations and masses,''
  Int.\ J.\ Mod.\ Phys.\ A {\bf 21}, 1901 (2006);
%
  A.~Bandyopadhyay,
S.~Choubey, S.~Goswami, S.~T.~Petcov and D.~P.~Roy,
%   ``Update Of The Solar Neutrino Oscillation Analysis With The 766-Ty  Kamland Spectrum,''
  Phys.\ Lett.\ B {\bf 608}, 115 (2005);
%  [arXiv:hep-ph/0406328].
%%CITATION = PHLTA,B608,115;%%
%
  G.~L.~Fogli {\it et al.},
%, E.~Lisi, A.~Marrone and A.~Palazzo,
  %``Global analysis of three-flavor neutrino masses and mixings,''
  Prog.\ Part.\ Nucl.\ Phys.\  {\bf 57}, 742 (2006).
%  [arXiv:hep-ph/0506083].
%%CITATION = PPNPD,57,742;%%

\bibitem{t2k}
  Y.~Itow {\it et al.},
  %``The JHF-Kamioka neutrino project,''
  arXiv:hep-ex/0106019.
  %%CITATION = HEP-EX 0106019;%%

\bibitem{nova}
  D.~S.~Ayres {\it et al.}  [NOvA Collaboration],
  %``NOvA proposal to build a 30-kiloton off-axis detector to study neutrino
  %oscillations in the Fermilab NuMI beamline,''
  arXiv:hep-ex/0503053.
  %%CITATION = HEP-EX 0503053;%%


\bibitem{reactor}
  F.~Ardellier {\it et al.},
  %``Letter of intent for double-CHOOZ: A search for the mixing angle
  %theta(13),''
  arXiv:hep-ex/0405032;
  %%CITATION = HEP-EX/0405032;%%
  X.~Guo {\it et al.}  [Daya-Bay Collaboration],
  %``A precision measurement of the neutrino mixing angle theta(13) using
  %reactor antineutrinos at Daya Bay,''
  arXiv:hep-ex/0701029.
  %%CITATION = HEP-EX/0701029;%%

\bibitem{iss}
http://www.hep.ph.ic.ac.uk/iss/

\bibitem{issphysics}
  A.~Bandyopadhyay {\it et al.}  [ISS Physics Working Group],
  %``Physics at a future Neutrino Factory and super-beam facility,''
  arXiv:0710.4947 [hep-ph].
  %%CITATION = ARXIV:0710.4947;%%

\bibitem{intrinsic}
  J.~Burguet-Castell,
M.~B.~Gavela, J.~J.~G\'{o}mez-Cadenas, P.~Hernandez and O.~Mena,
  %``On the measurement of leptonic CP violation,''
  Nucl.\ Phys.\ B {\bf 608}, 301 (2001).
%  [arXiv:hep-ph/0103258].
%%CITATION = NUPHA,B608,301;%%

\bibitem{minadeg}
  H.~Minakata and H.~Nunokawa,
  %``Exploring neutrino mixing with low energy superbeams,''
  JHEP {\bf 0110}, 001 (2001).
%  [arXiv:hep-ph/0108085].
%%CITATION = JHEPA,0110,001;%%

\bibitem{th23octant}
  G.~L.~Fogli and E.~Lisi,
%   ``Tests of three-flavor mixing in long-baseline neutrino oscillation
  %experiments,''
  Phys.\ Rev.\ D {\bf 54}, 3667 (1996).
%  [arXiv:hep-ph/9604415].
%%CITATION = PHRVA,D54,3667;%%

\bibitem{eight}
  V.~Barger, D.~Marfatia and K.~Whisnant,
%   ``Breaking eight-fold degeneracies in neutrino CP violation, mixing, and
  %mass hierarchy,''
  Phys.\ Rev.\ D {\bf 65}, 073023 (2002).
%  [arXiv:hep-ph/0112119].
%%CITATION = PHRVA,D65,073023;%%

\bibitem{zucc}
P.~Zucchelli,
%``A novel concept for a anti-nu/e / nu/e neutrino factory: The beta beam,''
Phys.\ Lett.\ B {\bf 532}, 166 (2002).
%%CITATION = PHLTA,B532,166;%%

\bibitem{volpe}
For a recent review see
  C.~Volpe,
  %``Topical review on 'beta-beams',''
  J.\ Phys.\ G {\bf 34}, R1 (2007).
%  [arXiv:hep-ph/0605033].

\bibitem{book_betabeam}
The book written by Mats Lindroos and Mauro Mezzetto: {\em BETA BEAMS}
(World Scientific, July, 2009).

%\cite{Agarwalla:2009xc}
\bibitem{agarwalla}
  S.~K.~Agarwalla,
    %``Some aspects of neutrino mixing and oscillations,''
      arXiv:0908.4267 [hep-ph].
        %%CITATION = ARXIV:0908.4267;%%

\bibitem{cernmemphys}
 J.~E.~Campagne {\it et al.},
 JHEP {\bf 0704}, 003 (2007).


\bibitem{paper1}
  S.~K.~Agarwalla, A.~Raychaudhuri and A.~Samanta,
%   ``Exploration prospects of a long baseline beta beam neutrino experiment
  %with an iron calorimeter detector,''
  Phys.\ Lett.\ B {\bf 629}, 33 (2005);
%  [arXiv:hep-ph/0505015].
  %%CITATION = HEP-PH 0505015;%%
  S.~K.~Agarwalla, S.~Choubey and A.~Raychaudhuri,
  %``Neutrino mass hierarchy and theta(13) with a magic baseline beta-beam
  %experiment,''
  Nucl.\ Phys.\  B {\bf 771}, 1 (2007).
%  [arXiv:hep-ph/0610333].
  %%CITATION = NUPHA,B771,1;%%

\bibitem{shortnote}
  S.~K.~Agarwalla, S.~Choubey and A.~Raychaudhuri,
  %``Unraveling neutrino parameters with a magical beta-beam experiment at
  %INO,''
  Nucl.\ Phys.\  B {\bf 798}, 124 (2008).
%  arXiv:0711.1459 [hep-ph].
  %%CITATION = ARXIV:0711.1459;%%

\bibitem{optimization}
  S.~K.~Agarwalla, S.~Choubey, A.~Raychaudhuri and W.~Winter,
  %``Optimizing the greenfield Beta-beam,''
  JHEP {\bf 0806}, 090 (2008).
%  arXiv:0802.3621 [hep-ex].
  %%CITATION = ARXIV:0802.3621;%%

\bibitem{pee}
   S.~K.~Agarwalla, S.~Choubey, S.~Goswami and A.~Raychaudhuri,
  %``Neutrino parameters from matter effects in $P_{ee}$ at long baselines,''
  Phys.\ Rev.\  D {\bf 75}, 097302 (2007).
%  [arXiv:hep-ph/0611233].
  %%CITATION = PHRVA,D75,097302;%%

\bibitem{two_baseline}
  S.~K.~Agarwalla, S.~Choubey and A.~Raychaudhuri,
  %``Exceptional Sensitivity to Neutrino Parameters with a Two Baseline
  %Beta-Beam Set-up,''
  Nucl.\ Phys.\  B {\bf 805}, 305 (2008).
%  [arXiv:0804.3007 [hep-ph]].
  %%CITATION = NUPHA,B805,305;%%

\bibitem{rparity}
  R.~Adhikari, S.~K.~Agarwalla and A.~Raychaudhuri,
%   ``Can R-parity violating supersymmetry be seen in long baseline beta-beam
  %experiments?,''
  Phys.\ Lett.\ B {\bf 642}, 111 (2006);
%  arXiv:hep-ph/0608034;
  %%CITATION = HEP-PH 0608034;%%
  S.~K.~Agarwalla, S.~Rakshit and A.~Raychaudhuri,
  %``Probing lepton number violating interactions with beta-beams,''
  Phys.\ Lett.\  B {\bf 647}, 380 (2007).
%  [arXiv:hep-ph/0609252].
  %%CITATION = PHLTA,B647,380;%%

%\cite{Agarwalla:2009em}
\bibitem{sanjib_vt1}
  S.~K.~Agarwalla, P.~Huber and J.~M.~Link,
  %``Constraining sterile neutrinos with a low energy beta-beam,''
  arXiv:0907.3145 [hep-ph].
  %%CITATION = ARXIV:0907.3145;%%

\bibitem{oldpapers}
  M.~Mezzetto,
  %``Physics reach of the beta beam,''
  J.\ Phys.\ G {\bf 29}, 1771 (2003);
%  [arXiv:hep-ex/0302007].
  %%CITATION = JPHGB,G29,1771;%%
%
  M.~Mezzetto,
  %``Beta beams,''
  Nucl.\ Phys.\ Proc.\ Suppl.\  {\bf 143}, 309 (2005);
%  [arXiv:hep-ex/0410083].
  %%CITATION = NUPHZ,143,309;%%
%
  M.~Mezzetto,
  %``Physics potential of the gamma = 100,100 beta beam,''
  Nucl.\ Phys.\ Proc.\ Suppl.\  {\bf 155}, 214 (2006).
%  [arXiv:hep-ex/0511005].
  %%CITATION = NUPHZ,155,214;%%
%

\bibitem{donini130}
  A.~Donini, E.~Fernandez-Martinez, P.~Migliozzi, S.~Rigolin and
L.~Scotto Lavina,
  %``Study of the eightfold degeneracy with a standard beta-beam and a
  %super-beam facility,''
  Nucl.\ Phys.\  B {\bf 710}, 402 (2005).
%  [arXiv:hep-ph/0406132].
  %%CITATION = NUPHA,B710,402;%%

\bibitem{doninibeta}
  A.~Donini, E.~Fernandez, P.~Migliozzi, S.~Rigolin, L.~Scotto Lavina,
  T.~Tabarelli de Fatis and F.~Terranova,
  %``Perspectives for a neutrino program based on the upgrades of the CERN
  %accelerator complex,''
  arXiv:hep-ph/0511134;
%%CITATION = HEP-PH/0511134;%%
%
  A.~Donini, E.~Fernandez-Martinez, P.~Migliozzi, S.~Rigolin,
  L.~Scotto Lavina, T.~Tabarelli de Fatis and F.~Terranova,
  %``A beta beam complex based on the machine upgrades of the LHC,''
  Eur.\ Phys.\ J.\  C {\bf 48}, 787 (2006).
%  [arXiv:hep-ph/0604229].
  %%CITATION = EPHJA,C48,787;%%

\bibitem{newdonini}
  P.~Coloma, A.~Donini, E.~Fernandez-Martinez and J.~Lopez-Pavon,
  %``$\theta_{13}$, $\delta$ and the neutrino mass hierarchy at a $\gamma=350$
  %double baseline Li/B $\beta$-Beam,''
  JHEP {\bf 0805}, 050 (2008);
%  [arXiv:0712.0796 [hep-ph]].
  %%CITATION = JHEPA,0805,050;%%
%\cite{Choubey:2009ks}
  S.~Choubey, P.~Coloma, A.~Donini and E.~Fernandez-Martinez,
  %``Optimized Two-Baseline Beta-Beam Experiment,''
  arXiv:0907.2379 [hep-ph].
  %%CITATION = ARXIV:0907.2379;%%

\bibitem{bc1}
  J.~Burguet-Castell, D.~Casper, E.~Couce, J.~J.~G\'{o}mez-Cadenas and P.~Hernandez,
  %``Optimal beta-beam at the CERN-SPS,''
  Nucl.\ Phys.\  B {\bf 725}, 306 (2005).
%  [arXiv:hep-ph/0503021].
  %%CITATION = NUPHA,B725,306;%%

\bibitem{bc2}
  J.~Burguet-Castell, D.~Casper, J.~J.~G\'{o}mez-Cadenas, P.~Hernandez and F.~Sanchez,
  %``Neutrino oscillation physics with a higher gamma beta-beam,''
  Nucl.\ Phys.\  B {\bf 695}, 217 (2004).
%  [arXiv:hep-ph/0312068].
  %%CITATION = NUPHA,B695,217;%%

\bibitem{fnal}
  A.~Jansson, O.~Mena, S.~J.~Parke and N.~Saoulidou,
  %``Combining CPT-conjugate Neutrino channels at Fermilab,''
  Phys.\ Rev.\  D {\bf 78}, 053002 (2008).
%  [arXiv:0711.1075 [hep-ph]].
  %%CITATION = PHRVA,D78,053002;%%

\bibitem{betaoptim}
  P.~Huber, M.~Lindner, M.~Rolinec and W.~Winter,
  %``Physics and optimization of beta-beams: From low to very high gamma,''
  Phys.\ Rev.\  D {\bf 73}, 053002 (2006).
%  [arXiv:hep-ph/0506237].
  %%CITATION = PHRVA,D73,053002;%%

\bibitem{doninialter}
  A.~Donini and E.~Fernandez-Martinez,
  %``Alternating ions in a beta-beam to solve degeneracies,''
  Phys.\ Lett.\ B {\bf 641}, 432 (2006).
%  [arXiv:hep-ph/0603261].
%%CITATION = PHLTA,B641,432;%%

%\cite{Meloni:2008it}
\bibitem{boulby}
  D.~Meloni, O.~Mena, C.~Orme, S.~Palomares-Ruiz and S.~Pascoli,
  %``An intermediate gamma beta-beam neutrino experiment with long baseline,''
  JHEP {\bf 0807}, 115 (2008).
%  [arXiv:0802.0255 [hep-ph]].
  %%CITATION = JHEPA,0807,115;%%

\bibitem{dusel}
http://www.lbl.gov/nsd/homestake/

\bibitem{duselwhite}
S.~Raby {\it et al.},
%``DUSEL Theory White Paper,''
arXiv:0810.4551 [hep-ph].
%%CITATION = ARXIV:0810.4551;%%

\bibitem{uslongbaseline}
V.~Barger {\it et al.},
%``Report of the US long baseline neutrino experiment study,''
arXiv:0705.4396 [hep-ph].
%%CITATION = ARXIV:0705.4396;%%

\bibitem{lindroos}
 M.~Lindroos,
  %``The acceleration and storage of radioactive ions for a beta-beam
  %facility,''
  arXiv:physics/0312042;
  %%CITATION = PHYSICS/0312042;%%
%
  M.~Lindroos,
  %``The Technical Challenges Of Beta-Beams,''
  Nucl.\ Phys.\ Proc.\ Suppl.\  {\bf 155}, 48 (2006).
  %%CITATION = NUPHZ,155,48;%%

\bibitem{betabeampage}
http://beta-beam.web.cern.ch/beta\%2Dbeam/

\bibitem{beta}
L. P. Ekstrom and R. B. Firestone, WWW Table of Radioactive Isotopes, \\
database version 2/28/99 from URL http://ie.lbl.gov/toi/

\bibitem{beamnorm}
  B.~Autin, R.~C.~Fernow, S.~Machida and D.~A.~Harris,
  %``NuFact02 machine working group summary,''
  J.\ Phys.\ G {\bf 29}, 1637 (2003);
  %%CITATION = JPHGB,G29,1637;%%
%
  F.~Terranova, A.~Marotta, P.~Migliozzi and M.~Spinetti,
  %``High energy beta beams without massive detectors,''
  Eur.\ Phys.\ J.\  C {\bf 38}, 69 (2004).
%  [arXiv:hep-ph/0405081].
  %%CITATION = EPHJA,C38,69;%%

\bibitem{prem}
  A.~M.~Dziewonski and D.~L.~Anderson,
  %``Preliminary Reference Earth Model,''
  Phys.\ Earth Planet.\ Interiors {\bf 25}, 297 (1981);
%%CITATION = PEPIA,25,297;%%
\\
S.~V.~Panasyuk, Reference Earth Model (REM) webpage,\\
 http://cfauves5.harvrd.edu/lana/rem/index.html.

\bibitem{msw1}
  L.~Wolfenstein,
  Phys.\ Rev.\ D {\bf 17}, 2369 (1978);
%%CITATION = PHRVA,D17,2369;%%

\bibitem{msw2}
  S.~P.~Mikheev and A.~Y.~Smirnov,
  Sov.\ J.\ Nucl.\ Phys.\  {\bf 42}, 913 (1985)
  [Yad.\ Fiz.\  {\bf 42}, 1441 (1985)];
%%CITATION = YAFIA,42,1441;%%
%
  S.~P.~Mikheev and A.~Y.~Smirnov,
  Nuovo Cim.\ C {\bf 9}, 17 (1986).
%%CITATION = NUCIA,9C,17;%%

\bibitem{msw3}
  V.~D.~Barger, K.~Whisnant, S.~Pakvasa and R.~J.~N.~Phillips,
  %``Matter effects on three-neutrino oscillations,''
  Phys.\ Rev.\ D {\bf 22}, 2718 (1980).
%%CITATION = PHRVA,D22,2718;%%

\bibitem{golden}
  A.~Cervera, A.~Donini, M.~B.~Gavela, J.~J.~G\'{o}mez-Cadenas, P.~Hernandez, O.~Mena and S.~Rigolin,
  %``Golden measurements at a neutrino factory,''
  Nucl.\ Phys.\ B {\bf 579}, 17 (2000)
  [Erratum-ibid.\ B {\bf 593}, 731 (2001)].
%  [arXiv:hep-ph/0002108].
  %%CITATION = HEP-PH 0002108;%%

\bibitem{freund}
  M.~Freund, P.~Huber and M.~Lindner,
  %``Systematic exploration of the neutrino factory parameter space  including
  %errors and correlations,''
  Nucl.\ Phys.\  B {\bf 615}, 331 (2001).
%  [arXiv:hep-ph/0105071].
  %%CITATION = NUPHA,B615,331;%%

\bibitem{ishihara}
  C.~Ishihara,
  %``How to do a $\nu_e \to \nu_\mu$ measurement in a SK-like detector,''
  PoS {\bf NUFACT08}, 044 (2008); For details, see the talk given by
Chizue Ishihara at EUROnu annual meeting at CERN, Geneva, 23-27th March, 2009.
%  [arXiv:0912.1002 [hep-ex]].
%%CITATION = POSCI,NUFACT08,044;%%

%\cite{Amerio:2004ze}
\bibitem{t600}
  S.~Amerio {\it et al.}  [ICARUS Collaboration],
  %``Design, construction and tests of the ICARUS T600 detector,''
  Nucl.\ Instrum.\ Meth.\  A {\bf 527}, 329 (2004).
  %%CITATION = NUIMA,A527,329;%%

\bibitem{bonnie}
 Bonnie T. Fleming, private communication.

\bibitem{wbb}
  V.~Barger, M.~Dierckxsens, M.~Diwan, P.~Huber, C.~Lewis, D.~Marfatia and B.~Viren,
  %``Precision physics with a wide band super neutrino beam,''
  Phys.\ Rev.\  D {\bf 74} (2006) 073004;
%  [arXiv:hep-ph/0607177];
  %%CITATION = PHRVA,D74,073004;%%
  M.~Diwan {\it et al.},
  %``Proposal for an experimental program in neutrino physics and proton  decay
  %in the homestake laboratory,''
  arXiv:hep-ex/0608023;
  %%CITATION = HEP-EX/0608023;%%
 V.~Barger, P.~Huber, D.~Marfatia and W.~Winter,
  %``Upgraded experiments with super neutrino beams: Reach versus Exposure,''
  Phys.\ Rev.\  D {\bf 76}, 031301 (2007);
%  [arXiv:hep-ph/0610301];
  %%CITATION = PHRVA,D76,031301;%%
V.~Barger, P.~Huber, D.~Marfatia and W.~Winter,
  %``Which long-baseline neutrino experiments are preferable?,''
  Phys.\ Rev.\  D {\bf 76}, 053005 (2007).
 % [arXiv:hep-ph/0703029].
  %%CITATION = PHRVA,D76,053005;%%


%\cite{Bishai:2008zza}
\bibitem{Bishai:2008zza}
  M.~Bishai,
  %``The US long baseline neutrino experiment study,''
  AIP Conf.\ Proc.\  {\bf 981}, 102 (2008).
  %%CITATION = APCPC,981,102;%%

\bibitem{fnalprotonplan}
http://www.fnal.gov/directorate/OPMO/Projects/PP/DirRev/2006/08\_15/review.htm

\bibitem{globes}

 P.~Huber, J.~Kopp, M.~Lindner, M.~Rolinec and W.~Winter,
  %``New features in the simulation of neutrino oscillation experiments with
  %GLoBES 3.0,''
  Comput.\ Phys.\ Commun.\  {\bf 177}, 432 (2007);
%  [arXiv:hep-ph/0701187]; %
  %%CITATION = CPHCB,177,432;%%
%
P.~Huber, M.~Lindner and W.~Winter,
  %``Simulation of long-baseline neutrino oscillation experiments with
  %GLoBES,''
  Comput.\ Phys.\ Commun.\  {\bf 167}, 195 (2005).
 % [arXiv:hep-ph/0407333].
  %%CITATION = CPHCB,167,195;%%


%\cite{Maltoni:2004ei}
\bibitem{Maltoni:2004ei}
  M.~Maltoni, T.~Schwetz, M.~A.~Tortola and J.~W.~F.~Valle,
  %``Status of global fits to neutrino oscillations,''
  New J.\ Phys.\  {\bf 6}, 122 (2004).
%  [arXiv:hep-ph/0405172].
  %%CITATION = NJOPF,6,122;

%\cite{Huber:2009cw}
\bibitem{Huber:2009cw}
  P.~Huber, M.~Lindner, T.~Schwetz and W.~Winter,
%``First hint for CP violation in neutrino oscillations from upcoming
%superbeam and reactor experiments,''
JHEP {\bf 0911}, 044 (2009).
%[arXiv:0907.1896 [hep-ph]].
%%CITATION = JHEPA,0911,044;%%


%\cite{Huber:2002mx}
\bibitem{Huber:2002mx}
  P.~Huber, M.~Lindner and W.~Winter,
%``Superbeams versus neutrino factories,''
  Nucl.\ Phys.\  B {\bf 645}, 3 (2002).
%[arXiv:hep-ph/0204352].
%%CITATION = NUPHA,B645,3;%%


\end{thebibliography}
\end{document}